\documentclass[10pt,amsmath,amssymb,aps,pra,showpacs,a4paper,twocolumn,superscriptaddress]{revtex4-1}

\usepackage[cp1250]{inputenc} 

\usepackage{dsfont, color}
\usepackage{braket}
\usepackage{enumerate}
\usepackage{graphicx, epsfig}
\usepackage{lmodern}
\usepackage{mathptmx}

\def\beq{\begin{equation}}
\def\eeq{\end{equation}}
\def\beqa{\begin{eqnarray}}
\def\eeqa{\end{eqnarray}}

\newcommand{\eq}[1]{Eq.(#1)}
\newcommand{\fig}[1]{Fig.(#1)}

\begin{document}

\title{Dirac fields in curved spacetime as Fermi-Hubbard model with non unitary tunnelings}

\author{Ji\v{r}í Miná\v{r}}
\affiliation{Centre for Quantum Technologies, National University of Singapore, 3 Science Drive 2, Singapore 117543, Singapore}
\author{Beno\^{\i}t~Gr\'{e}maud}
\affiliation{Centre for Quantum Technologies, National University of Singapore, 3 Science Drive 2, Singapore 117543, Singapore}
\affiliation{Department of Physics, National University of Singapore, 2 Science Drive 3, Singapore 117542, Singapore}
\affiliation{ Merlion MajuLab, CNRS-UNS-NUS-NTU International Joint Research Unit UMI 3654, Singapore  }
\affiliation{Laboratoire Kastler Brossel, Ecole Normale Sup\'{e}rieure CNRS, UPMC; 4 Place Jussieu, 75005 Paris, France}



\begin{abstract}
	In this article we show that a Dirac Hamiltonian in a curved background spacetime can be interpreted, when discretized, as a tight binding Fermi-Hubbard model with non unitary tunnelings. We find the form of the nonunitary tunneling matrices in terms of the metric tensor. The main motivation behind this exercise is the feasibility of such Hamiltonians by means of laser assisted tunnelings in cold atomic experiments. The mapping thus provide a physical interpretation of such Hamiltonians. We demonstrate the use of the mapping	on the example of time dependent metric in 2+1 dimensions. Studying the spin dynamics, we find qualitative agreement with known theoretical predictions, namely the particle pair creation in expanding universe.
\end{abstract}

\maketitle

\section{Introduction}
\label{sec Introduction}

Recently, much interest was devoted to a study of many body physics of quantum gases \cite{Ketterle_2008, Bloch_2008}. High degree of control of the experimental parameters has allowed for designing specific Hamiltonians \cite{Lewenstein_2007}. A special category is then a fabrication of synthetic gauge fields \cite{Dalibard_2011}, where a remarkable experimental progress has been achieved in last couple of years, including the realization of synthetic electric \cite{Lin_2011} and magnetic \cite{Lin_2009} fields in the bulk as well as on the lattice \cite{Aidelsburger_2011}. A non abelian synthetic gauge field of the spin orbit Rashba type has been demonstrated in the bulk \cite{Lin_2011b, ZhangJY_2012}. In the case of a lattice, which will be of main interest, laser assisted hoppings \cite{Jaksch_2003, Gerbier_2010} allow for a simulation of a (non abelian) lattice gauge theory \cite{Kogut_1979} with cold atoms \cite{Osterloh_2005}. Different works adressed the question of non abelian background fields with cold atoms. Such situation occurs e.g. in the case of electrons with spin orbit coupling and can be studied with cold atomic systems, in both non interacting \cite{Goldman_2009} and interacting \cite{Cocks_2012, Cole_2012} cases. In those scenarios, however, the tunnelings of different spin components between two adjacent lattice sites are described by unitary matrices, due to the hermiticity of the gauge field \cite{Kogut_1979, Makeenko_2002}. Moreover, an explicit form of these matrices is determined by the theory one wants to simulate, e.g. the mentioned spin orbit coupling. It is thus an interesting question, what happens, if these tunnelings become non unitary, which can be, in principle, done in the cold atomic experiments using known techniques, as explained below.

The article is structured as follows. In Sec. \ref{sec Theoretical model} we derive the mapping between continuous Dirac fields in curved background spacetime and the Fermi-Hubbard model with general tunnelings (we also derive the non-relativistic limit of the mapping in the Appendix). In Sec. \ref{sec Expanding universe} we demonstrate the use of the mapping on a specific example of the expanding universe in 2+1 dimensions. We discuss in detail the dispersion relations in continuum and on the lattice and how they are related using adiabaticity criteria. In Sec. \ref{sec Implementations} we discuss how such systems can be implemented with cold atoms. We show how the toy example of expanding universe can be observed through the spin dynamics. Finally we conclude in Sec. \ref{sec Conclusion}.

In a simple case of a static diagonal metric, the tunnelings become unitary. Alternative interpretation of the Fermi-Hubbard Hamiltonian is that of a Pauli Hamiltonian, i.e. a non relativistic limit of the Dirac Hamiltonian. In this case the tunnelings remain, in general, non unitary even for the static diagonal metric.

\section{Theoretical model}
\label{sec Theoretical model}

\subsection{Continuum - lattice mapping}

Driven by this motivation, let us start with a kinetic fermionic Hamiltonian in $d-1$ spatial dimensions of a form
\beq
	H(t) = \sum_{{\bf x}, k, s, s'} \Psi^\dag_s({\bf x})T_{s,s'}(x,x+a_k) \Psi_{s'}({\bf x}+a_k) + {\rm h.c.},
	\label{eq H kin latt}
\eeq
where the sum runs over all lattice sites ${\bf x}$ and directions $k=1..d-1$ and the fermionic operators satisfy the usual anticommutation relations
\beq
	\{\Psi_s({\bf x}), \Psi^\dag_{s'}({\bf x'}) \} = \delta_{s,s'} \delta_{\bf x, x'}.
	\label{eq acomm rel psi}
\eeq
Throughout the paper, we denote the spacetime coordinates as $x$ $(=(t,{\bf x}))$, while the space coordinates as $\bf x$. The matrices $T(x,x+a_k)$ represent a parallel transporter of a quantum field $\psi$ between sites $x$ and $x + a_k$. In the case of lattice gauge theories, the matrix $T$ belongs to a representation of a gauge group, which is typically compact, such as $U(n)$ with $n$ being the number of "flavor" components of the field $\psi$. In such case, the matrices $T$ become unitary. The relevant question is thus, what if the $T$ belong to a non compact gauge group? This question has actually been already adressed in the past \cite{Cahill_1979, Cahill_1983, Wu_1975} and more recently \cite{Lehmann_2005}, but such interpretation seems to be problematic and was not actively pursued. 
Lets return to the Hamiltonian \eq{\ref{eq H kin latt}}. In what follows, we will be interested in physics of fermions, so that $\psi$ is a spinor. As discussed later, a general matrix $T \in Gl(n,\mathbb{C})$ can be engineered in cold atomic systems ($n$ is the number of spin components). We would like to emphasize, that for such a novel situation, the Hamiltonian \eq{\ref{eq H kin latt}} is interesting on its own right. However, it is also interesting to look, whether some physical significance can be given to it.

A starting point of our discussion will be a classical (not quantized) fermionic field in a curved spacetime, which can be described by a Lagrangian density \cite{Birell_1982, Kibble_1961}
\beq
	\mathcal{L}(x) = \sqrt{g} \left\{ \frac{1}{2}i \bar{\psi}(x) \underline{\gamma}^\mu D_\mu \psi(x) + {\rm h.c.} - m\bar{\psi}(x) \psi(x) \right\}.	
	\label{eq L}
\eeq
Let us recall \cite{Weinberg_1972}, that working in a coordinate basis $e_\mu$, in which the spacetime vector $x$ is defined in terms of its components $x^\mu$, $x = x^\mu e_\mu$, one may construct a local orthonormal basis $e_\alpha$. The two bases are related through vielbeins $e_\alpha = e^\mu_\alpha e_\mu$. The metric tensor $g$ is defined as $\eta_{\alpha \beta} = e^\mu_\alpha e^\nu_\beta g_{\mu \nu}$, where $\eta$ is the Minkowski, i.e. flat metric. Then, the curved spacetime $\gamma$ matrices are defined as \cite{Parker_2009} $\underline{\gamma}^\mu = \gamma^\alpha e^\mu_\alpha$, where $\gamma^\alpha$ are the usual (flat spacetime) Dirac matrices, for which $\{\gamma^\alpha, \gamma^\beta\} = 2 \eta^{\alpha \beta}$ and we adopt the sign convention $\eta = (+,-,-,...)$. The covariant derivative acting on the spinor is $D_\mu = \partial_\mu  - \Gamma_\mu$, where $\Gamma_\mu(x) = \frac{1}{8}\left[\gamma^\alpha, \gamma^\beta \right] e^\nu_\alpha (\nabla_\mu e_{\beta \nu})$ and $\nabla_\mu e_{\beta \nu} = \partial_\mu e_{\beta \nu} - \Gamma^\sigma_{\mu \nu} e_{\beta \sigma}$ (for a brief overview with essential technical details, see \cite{Yepez_2011}). The canonically conjugate momentum to $\psi$ can be found in a usual way as
\beq
	\pi(x) = \frac{\partial \mathcal{L}}{\partial(\partial_0 \psi)} = \sqrt{g} \frac{1}{2}i \bar{\psi} \underline{\gamma}^0
	\label{eq pi}
\eeq 
and similarly for $\bar{\pi}$ which is conjugate to $\bar{\psi}$. One then obtains the Hamiltonian density $\mathcal{H} = \pi (\partial_0 \psi) + (\partial_0 \bar{\psi}) \bar{\pi} - \mathcal{L}$,
\beq
	\mathcal{H}(x) =  - \sqrt{g} \left\{   \frac{1}{2}i \bar{\psi} \left[  \underline{\gamma}^k D_k -  \underline{\gamma}^0 \Gamma_0 \right] \psi + {\rm h.c.}  - m\bar{\psi} \psi \right\}. 
	\label{eq H dens}
\eeq
Lets now cosider an isotropic square lattice in coordinate basis with lattice spacing $a$. We introduce a covariant derivative on the lattice as
\beq
	D_\mu \psi(x) = \frac{1}{a}\left[ P(x,x + a_\mu) \psi(x + a_\mu) - \psi(x) \right],
\eeq
where $P(x,x + a_k)$ is the parallel transporter from $x + a_k$ to $x$ and reads (here $\mu$ is fixed)
\beq
	P(x,x + a_\mu) = \mathcal{P} {\rm exp} \left[ \int_{x + a_\mu}^{x} {\rm d}x^\mu \Gamma_\mu(x) \right]
	\label{eq par prop}
\eeq
and $\mathcal{P}$ stands for the path ordering. One can then formally discretize the Hamiltonian $H(t) = \int {\rm d}^{d-1} x \mathcal{H}(x)$,
\beqa
	H(t) &=& \frac{1}{2a} \sum_{{\bf x}, k} \psi^\dag(x) M^k P(x,x+a_k) \psi(x+a_k) \nonumber \\
	&& \,\, \psi^\dag(x) \left[ aM^0\Gamma_0 + \frac{1}{2}\sqrt{g}m\gamma^0 \right] \psi(x) + {\rm h.c.},
	\label{eq H Dirac latt}
\eeqa
where $M^\mu(x) \equiv -i \sqrt{g} \gamma^0 \underline{\gamma}^\mu$. At first sight, the structure of the Hamiltonian \eq{\ref{eq H Dirac latt}} is similar to \eq{\ref{eq H kin latt}}, but there are two major differences. First, in the former case, the fields are not quantized and second, they are time dependent, so it is not obvious, whether one can obtain the desired anticommutation relation \eq{\ref{eq acomm rel psi}} for the space dependent operators. In order to proceed, we shall rely on the arguments exposed in \cite{Huang_2009}. We summarize the main steps crucial for our purpose. The classical field $\psi(x)$ is a projection of ket $\ket{\psi}$ to a spatiotemporal basis $\ket{x}$
\beq
	\psi(x) = \braket{x|\psi}.
	\label{eq psi xt}
\eeq
Similarly, one obtains a different field which is only space dependent on some constant time hypersurface as 
\beq
	\Psi({\bf x}) = \braket{{\bf x}| \psi}. 
	\label{eq psi x}
\eeq
The relationship between the two fields $\psi(x)$ and $\Psi({\bf x})$ can be found from the equivalence $\braket{\phi | \psi} = (\phi, \psi)$, where the scalar product in curved spacetime is defined as \cite{Huang_2009}
\beq
	(\phi, \psi) = \int {\rm d}^{d-1}x \sqrt{g} \phi^\dag \gamma^0 \underline{\gamma}^0 \psi.
	\label{eq scal prod}
\eeq
From the resolution of identity $\int {\rm d}^{d-1} x \ket{x} \sqrt{g} \gamma^0 \underline{\gamma}^0 \bra{x} = \mathds{1}$ one obtains the evolution of the ket $\ket{x}$
\beq
	\partial_0 \ket{x} = -\frac{1}{2} \ket{x} (\partial_0 \sqrt{g} \gamma^0 \underline{\gamma}^0 ) ( \sqrt{g} \gamma^0 \underline{\gamma}^0 )^{-1},
\eeq
which can be formally integrated to give
\beq
	\ket{x} = \ket{t,{\bf x}} = \sqrt{2} \ket{\bf x} ( \sqrt{g} \gamma^0 \underline{\gamma}^0 )^{-\frac{1}{2}}.
	\label{eq ket xt}
\eeq
The factor $\sqrt{2}$ is an integration constant coming from the hermitian definition of the Lagrangian \eq{\ref{eq L}}.

The quantization of the space and time dependent field \eq{\ref{eq psi xt}}, $\psi(x)$, then proceeds by imposing equal time anticommutation relations for the canonically conjugate operators \cite{Parker_2009}, namely
\beq
	\{\psi_s(x), \pi_{s'}(x') \} = i \delta({\bf x - x}') \delta_{s, s'},
	\label{eq anticomm psi pi}
\eeq
where $\pi(x)$ is given by \eq{\ref{eq pi}}. We can use the relations \eq{\ref{eq psi xt}, \ref{eq psi x}} and \eq{\ref{eq ket xt}} to find the relationship between the two fields $\psi(x)$ and $\Psi({\bf x})$ to be
\beq
	\psi(x) = \sqrt{2} \left( \sqrt{g} \gamma^0 \underline{\gamma}^0 \right)^{-\frac{1}{2}} \Psi({\bf x}).
	\label{eq psi xt psi x} 
\eeq
Plugging \eq{\ref{eq pi}} into \eq{\ref{eq anticomm psi pi}} and using \eq{\ref{eq psi xt psi x}} we obtain for the anticommutator of the constant time hypersurface fields
\[
	\{\Psi_s({\bf x}), \Psi^\dag_{s'}({\bf x}') \} = \delta({\bf x - x}') \delta_{s, s'},
\]
which is precisely the relation \eq{\ref{eq acomm rel psi}}.

Lets now take the lattice Dirac Hamiltonian \eq{\ref{eq H Dirac latt}} and write it as
\beq
	H(t) = \sum_{{\bf x}, k} \psi^\dag(x) \tilde{T}(x, x + a_k) \psi(x + a_k) + {\rm h.c.} + \psi^\dag(x) \tilde{V}(x) \psi(x),	
	\label{eq H t temp}
\eeq
We now use the relation \eq{\ref{eq psi xt psi x}} to substitute for $\psi(x)$ and write the Hamiltonian as
\beqa
	&& H(t) = \nonumber \\
	&& \sum_{{\bf x}, k} \Psi^\dag({\bf x}) T(x, x + a_k) \Psi({\bf x} + a_k) + {\rm h.c.} + \Psi^\dag({\bf x}) V(x) \Psi({\bf x}),	
	\label{eq H t}
\eeqa
where
\beqa
	&& T(x,x+a_i) = \nonumber \\
	&& \frac{1}{2} (iM^0)^{-1/2}(x) M^i P(x,x+a_i) ((iM^0)^\dag)^{-1/2}(x+a_i) \nonumber \\
	&& V(x) = \nonumber \\
	&& (iM^0)^{-1/2} \left[ M^0\Gamma_0 + (M^0\Gamma_0)^\dag + \sqrt{g} m\gamma^0 \right] ((iM^0)^\dag)^{-1/2}, \nonumber \\
	&& 
	\label{eq T V}
\eeqa
where we have put the lattice spacing $a=1$ for simplicity. Now, the Hamiltonian \eq{\ref{eq H t}} has the same structure with the correct anticommutation relations for the operators as \eq{\ref{eq H kin latt}} (plus the local term). The price to pay in order to achieve this goal was to absorb the spatiotemporal dependence of the fields $\psi(x)$ to the elements of the Hamiltonian and thus spoiling its covariance.

\emph{Fermion doubling.} Due to the naive discretization of the Hamiltonian \eq{\ref{eq H dens}} one obtains the lattice formulation with doublers. Since in the present work we are interested only in noninteracting theory, and taking into account the fact, that one can address experimentally individual $\bf k$ vectors (cf. below), we don't elaborate on this issue further. Proposals how to deal with the fermion doubling in cold atomic experiments exist in the literature, see e.g. \cite{Banerjee_2013} for staggered fermions formulation.

\subsection{Physical interpretation}

We just provided a possible interpretation of the kinetic Hamiltonian of the form \eq{\ref{eq H t}} with general non unitary tunnelings $T$. In Appendix, we provide an alternative mapping, corresponding to the nonrelativistic limit of the Dirac Hamiltonian leading to the Pauli Hamiltonian, yielding formally the same expression for the lattice Hamiltonian \eq{\ref{eq H t}}.

The question is then what field theory is actually simulated. Lets take an example of simulation, where the Hamiltonian \eq{\ref{eq H kin latt}} describes a motion in a (two dimensional) plane for a two component field $\psi_s$, $s=1,2$. If we want to interpret it as a Dirac Hamiltonian, our simulator would correspond to a Dirac Hamiltonian in 2+1 dimensions. If we are to interpret it, however, as a Pauli Hamiltonian, the simulator corresponds to a spin half fermion living in 3+1 dimensions, but whose motion is confined to a plane.

We would like to mention, that a simulation of a Dirac field in curved spacetime with cold atoms was already adressed \cite{Boada_2011}, but the discretization was carried out in the limit of small lattice spacing, such that the approximation $P \approx 1 + a \Gamma$ is valid. For stationary metrics, considered in \cite{Boada_2011}, it results in unitary tunnelings.

\section{Case study: Expanding FLRW universe}
\label{sec Expanding universe}

In order to show the above derived mapping on some physically relevant scenario, we choose a textbook example of an expanding universe described by a Friedmann-Lema\^{i}tre-Robertson-Walker (FLRW) metric in 2+1 dimensions with a line element
\beq
	{\rm d}s^2 = {\rm d}t^2 - b_x(t)^2 {\rm d}x^2 - b_y(t)^2 {\rm d}y^2.
	\label{eq line el}
\eeq

Using the relations \eq{\ref{eq T V}} leads to the result
\beqa
	T_x &\equiv& T(x,x+a_x) = -\frac{i}{2b_x}\left[ \mathds{1}_2 \sinh \partial_0 b_x + \sigma_x \cosh \partial_0 b_x \right] \nonumber \\
	T_y &\equiv& T(x,x+a_y) = -\frac{i}{2b_y}\left[ \mathds{1}_2 \sinh \partial_0 b_y - \sigma_y \cosh \partial_0 b_y \right] \nonumber \\
	V(x) &=& m \sigma_z.
	\label{eq T V in FW}
\eeqa
It then clear, that in the lattice formulation the FLRW metric maps to a generalized time dependent spin-orbit coupling.

\subsection{Continuum dispersion}

Lets consider the continuum Hamiltonian density \eq{\ref{eq HD dens}}
\beq
	\mathcal{H}_D = (\underline{\gamma}^0)^{-1} (m - i\underline{\gamma}^k D_k) + i\Gamma_0. \nonumber
\eeq
One can derive from the line element \eq{\ref{eq line el}} the following relations
\beqa
	\sqrt{g} = b_x b_y & \nonumber \\
	\underline{\gamma}^0 = \gamma^0 & \Gamma^0 = 0 \nonumber \\
	\underline{\gamma}^1 = \frac{1}{b_x}\gamma^1 \phantom{cau} & \Gamma^1 = \left[\gamma^0,\gamma^1 \right] w_{011} = 2 \sigma_x \partial_0 b_x \nonumber \\
	\underline{\gamma}^2 = \frac{1}{b_y}\gamma^2 \phantom{cau} & \Gamma^2 = \left[\gamma^0,\gamma^2 \right] w_{022} = -2 \sigma_y \partial_0 b_y
\eeqa
Using the Dirac representation of $\gamma$ matrices $\gamma^0 = \sigma_z$, $\gamma^1 = i \sigma_y$, $\gamma^2 = i \sigma_x$ and the above relations, one gets
\beq
	\mathcal{H}_D = \gamma^0 m - \frac{i}{b_x} \left(\sigma_x \partial_x - 2 (\partial_0 b_x) \mathds{1}_2 \right) - \frac{i}{b_y} \left(-\sigma_y \partial_y + 2 (\partial_0 b_y) \mathds{1}_2 \right).
	\label{eq HD dens FW}
\eeq
Next, going to the Fourier space
\beq
	\psi(t,{\bf x}) = \frac{1}{2\pi} \int {\rm d}^2 k {\rm e}^{-i{\bf k} \cdot {\bf x}} \psi(t,{\bf k})
	\label{eq FT}
\eeq
the final Hamiltonian takes the form $H(t) = \int {\rm d}^2 k \mathcal{H}$, where
\beqa
	\mathcal{H} &=& \psi(t,{\bf k})^\dag \frac{1}{2} \left( \sqrt{g}\mathcal{H}_D + {\rm h.c.} \right) \psi(t,{\bf k}) \nonumber \\
	&=& \psi(t,{\bf k})^\dag \sqrt{g} \left( \begin{array}{cc}
m & -\frac{k_x}{b_x} - \frac{i k_y}{b_y} \\
-\frac{k_x}{b_x} + \frac{i k_y}{b_y} & -m \end{array} \right) \psi(t,{\bf k}) \nonumber \\
	&=& \Psi({\bf k})^\dag \left( \begin{array}{cc}
m & -\frac{k_x}{b_x} - \frac{i k_y}{b_y} \\
-\frac{k_x}{b_x} + \frac{i k_y}{b_y} & -m \end{array} \right) \Psi({\bf k}),
\eeqa
where in the last equality we have used the relation \eq{\ref{eq  psi xt psi x}} between the two fields $\psi(t,{\bf k}) = \sqrt{2} (\sqrt{g} \gamma^0 \underline{\gamma}^0)^{-\frac{1}{2}} \Psi({\bf k}) = \sqrt{2}/\sqrt{b_x b_y} \Psi({\bf k})$. We thus obtain the instantaneous eigenvalues of the total Hamiltonian in continuum (this result is compatible with the one of Ref. \cite{Huang_2009})
\beq
	\epsilon_\pm = \pm \sqrt{m^2 + \left( \frac{k_x}{b_x} \right)^2 + \left( \frac{k_y}{b_y} \right)^2}
	\label{eq disp cont}
\eeq

\subsection{Lattice dispersion}

Combining \eq{\ref{eq H t}}, \eq{\ref{eq T V in FW}} and the lattice version of \eq{\ref{eq FT}}, one gets for the lattice Hamiltonian
\beq
	H(t) = \sum_{k_x,k_y} \Psi^\dag_{\bf k} \left( \begin{array}{cc}
m - \mu_{\bf k} & -\gamma_{\bf k} \\
-\gamma^*_{\bf k} & -m - \mu_{\bf k} \end{array} \right) \Psi_{\bf k} ,
\label{eq Hk latt}
\eeq
where the coefficients read
\beqa
	\mu_{\bf k} &=& \frac{\sin{k_x}\sinh{\partial_0 b_x}}{b_x} + \frac{\sin{k_y}\sinh{\partial_0 b_y}}{b_y} \nonumber \\
	\gamma_{\bf k} &=& \frac{\sin{k_x}\cosh{\partial_0 b_x}}{b_x} + i\frac{\sin{k_y}\cosh{\partial_0 b_y}}{b_y}.
	\label{eq mu gamma}
\eeqa
The instantaneous eigenenergies on the lattice are
\beq
	\epsilon_\pm = - \mu_{\bf k} \pm \sqrt{m^2 + |\gamma_{\bf k}|^2}.
	\label{eq disp latt}
\eeq
One can also verify, that the lattice dispersion relation \eq{\ref{eq disp latt}} yields the dispersion relation \eq{\ref{eq disp cont}} in the continuum limit. So far we have been working with $\hbar = c = a = 1$. In order to find the continuum limit, we need to restore the lattice spacing $a$ in the equations. Noting, that the lattice spacing enter the lattice version of the Hamiltonian through the definition of the covariant derivative only ($\psi_x^\dag D_j \psi_x = \psi_x^\dag 1/a \left[ P_j \psi_{x+j} - \psi_x \right]$ with $P_j = {\rm exp}(-\Gamma_j a)$). In our case it translates to the multiplication of tunneling matrices $T$, \eq{\ref{eq T V in FW}}, by a factor $1/a$ ($T \rightarrow \frac{1}{a} T$) and since $\Gamma_j \propto \partial_0 b_j$, by multiplication of the occurences of $\partial_0 b_j$ by $a$ ($\partial_0 b_j \rightarrow a \partial_0 b_j$). Written explicitly, the coefficients $\mu_{\bf k}, \gamma_{\bf k}$ (\eq{\ref{eq mu gamma}}) become
\beqa
	\mu_{\bf k} &=& \frac{\sin{a k_x}\sinh{a \partial_0 b_x}}{a b_x} + \frac{\sin{a k_y}\sinh{a \partial_0 b_y}}{a b_y} \nonumber \\
	\gamma_{\bf k} &=& \frac{\sin{a k_x}\cosh{a \partial_0 b_x}}{a b_x} + i\frac{\sin{a k_y}\cosh{a \partial_0 b_y}}{a b_y},
\eeqa
which in the continuum limit gives
\beqa
	\lim_{a \to 0^+} \mu_{\bf k} &=& 0 \nonumber \\
	\lim_{a \to 0^+} \gamma_{\bf k} &=&	\frac{k_x}{b_x} + i \frac{k_y}{b_y},
\eeqa
which in turn yields the continuum dispersion relation as expected.

\subsection{Note on adiabaticity}
\label{subsec Note on adiabaticity}

We have shown, that in the limit $a \rightarrow 0^+$, one recovers the correct continuum dispersion relation. In practice, however, the lattice spacing $a$ is finite and thus care must be taken when interpreting the results of the simulation in terms of continuum theory. Intuitively, one expects to recover the continuum theory in the limit of small wavevectors $k$, since the details of the underlying lattice should not be important. In our specific example we thus consider the limit ($a$ nonzero) $ak = \frac{2\pi}{k_{\rm max}}k \ll 1$. In this limit, the dispersion relation \eq{\ref{eq disp latt}} becomes
\beq
	\epsilon_\pm \approx \frac{ak_x\,a\partial_0 b_x}{ab_x} + \frac{ak_y\,a\partial_0 b_y}{ab_y} \pm \sqrt{m^2 + \left( \frac{ak_x}{ab_x}\right)^2 + \left( \frac{ak_y}{ab_y}\right)^2},
	\label{eq disp latt approx}
\eeq
where we assume sufficiently slow changes in $b$, such that only leading term in the expansions of functions $\sinh (a\partial_0 b)$, $\cosh (a \partial_0 b)$ is dominant. In order to recover the continuum dispersion, the $\sqrt{\phantom{cau}}$ term has to be dominant. Explicitly, we can consider two limiting cases (i) $m \approx 0 \ll k_{j}/b_{j}$ and (ii) $m \gg k_{j}/b_{j}$. In these two cases we thus have
\beqa
	m \ll \frac{k_{j}}{b_{j}} &\Rightarrow& \frac{k_{j}a \partial_0 b_{j}}{b_{j}} \ll \frac{k_{j}}{b_{j}} \Rightarrow \boxed{a\partial_0 b_{j} \ll 1} \nonumber \\
	\frac{k_{j}}{b_{j}} \ll m &\Rightarrow& \frac{k_{j}a \partial_0 b_{j}}{b_{j}} \ll m \Rightarrow \boxed{a\partial_0 b_{j} \approx 1},
	\label{eq ad criteria}
\eeqa
where $j=x,y$. These conditions can be once again understood intuitively such that the characteristic rate $\omega$ of change of $b$ has to be much smaller than the maximum frequency supported by the lattice $\omega_{\rm max} = k_{\rm max} = \frac{2\pi}{a}$ (with $c=1$).

An important remark to make here, which is relevant for the implementations with cold atoms, is that a particularly important case is half filling. In the massless case, the dispersion relation \eq{\ref{eq disp latt}} yields Dirac cones at $k_{x,y} = 0, \pm \pi$ with Fermi energy $\epsilon_{\rm F}=0$. The development around the Dirac points ${\bf k} \rightarrow {\bf k}_{\rm Dirac} + {\bf k}$ yields the dispersion relation for small $k$ \eq{\ref{eq disp latt approx}}. In other words, it is natural to work at half filling, where the relevant wavevectors lie in the vicinity of the Dirac points, which approximates well the continuum theory.

\section{Implementation with cold atoms}
\label{sec Implementations}

In the experiments with cold atoms, the internal degrees of freedom are usually played by the hyperfine states of the atoms \cite{Bloch_2008}. These allow for a laser assisted tunnelings between adjacent sites, say $i,i'$ of the optical lattice. Lets denote the internal degrees of freedom $s$. In order to engineer an arbitrary $T(x) \in Gl(n,\mathbb{C})$, it is necessary to control each of the tunneling rates $(i,s) \leftrightarrow (i',s')$ independently in all spatial directions and moreover, the rates in general vary in spacetime. Different techniques and their combination can be used in order to achieve this goal. For example, bichromatic lattices can be combined with an independent Raman laser for each transition $s \leftrightarrow s'$ \cite{Mazza_2012}. The spatial dependence is then given by a transverse profile of each Raman laser. It can be given e.g. by a (typically) gaussian laser profile which varies slowly on the lattice spacing or it can be designed using a specific phase masks \cite{Bakr_2009} or array of microlenses \cite{Itah_2010}, which allow for the modulation on the scale of lattice spacing and were already used in the cold atomic experiments. Another comment is, that the potential $V(x)$ in \eq{\ref{eq H t}} is non diagonal and might be difficult to engineer. The way around is that since $V$ is hermitian, it can be diagonalized by unitary transformation. It amounts to redefine the tunneling matrix $T$ (analogous to a local gauge transformation in the case of gauge fields) and the spinors $\Psi$. Since the transformation is unitary, the anticommutation relations \eq{\ref{eq acomm rel psi}} are preserved.

\subsection{Expanding FLRW universe: Time evolution of the spins}

In this section we will investigate the dynamics of a typical observable accessible with cold atomic systems. We would like to emphasize, that we are interested only in qualitative features of our mapping with respect to the continuum theory. More formal and quantitative comparison between the continuum theory and its lattice counterpart is in principle possible (using the analysis in \cite{Parker_1971, Parker_1989, Birell_1982}), however it requires a significant amount of additional work, which is not crucial for the conclusions presented in the following.

We consider the spin $S$ defined either in the physical spin space (spanned by operators $\Psi_{{\bf k},s}, \Psi_{{\bf k},s}^\dag$) or the spin $\mathcal{S}$ defined in the local diagonal basis spanned by operators $d_{{\bf k},s}, d_{{\bf k},s}^\dag$. They are defined as
\beqa
	S^a_{\bf k}(t) &=& \frac{1}{2} \braket{\Psi_{{\bf k},s}^\dag(t) \sigma^a_{ss'} \Psi_{{\bf k},s'}(t) }  \\
	\mathcal{S}^a_{\bf k}(t) &=& \frac{1}{2} \braket{d_{{\bf k},s}^\dag(t) \sigma^a_{ss'} d_{{\bf k},s'}(t) },
	\label{eq spins}
\eeqa
where $\sigma$ are the usual Pauli matrices. Working in the Heisenberg picture, an evolution of an arbitrary operator $O$ is governed by the Heisenberg equations of motion $\dot{O} = -i \left[ O, H \right]$ with the Hamiltonian \eq{\ref{eq Hk latt}}. Written in components the equation of motion reads
\beq
	\dot{O}_{{\bf k},s} = -i H_{{\bf k},ss'} O_{{\bf k},s'}.
	\label{eq dot O}
\eeq
Defining spinors as $\Psi_{\bf k} = (\Psi_{{\bf k},1},\Psi_{{\bf k},2})^{\rm T}$ and $d_{\bf k} = (d_{{\bf k},+},d_{{\bf k},-})^{\rm T}$ one can introduce the relation between the two bases as
\beq
	d_{\bf k}(t) = U_{\bf k}^\dag(t) \Psi_{\bf k}(t)
\eeq
and the time evolution operator as
\beq
	\Psi_{\bf k}(t) = \tilde{K}_{\bf k}(t,t_0) U_{\bf k}(t_0) d_{\bf k}(t_0) \equiv K_{\bf k}(t,t_0) d_{\bf k}(t_0).
\eeq
Plugging these relations to the definitions of the spin observables \eq{\ref{eq spins}}, one gets
\beqa
	S^a_{\bf k}(t) &=& \frac{1}{2} \braket{d_{{\bf k},s}^\dag(t_0) K^\dag_{{\bf k},sr}(t,t_0) \sigma^a_{rr'} K_{{\bf k},r's'}(t,t_0) d_{{\bf k},s'}(t_0) }  \nonumber \\
	&=& \frac{1}{2} K^\dag_{{\bf k},sr}(t,t_0) \sigma^a_{rr'} K_{{\bf k},r's'}(t,t_0) n_{{\bf k},s}(t_0) \delta_{ss'}  \nonumber \\
	&=& \frac{1}{2} {\rm Tr} \left( K^\dag_{{\bf k}}(t,t_0) \sigma^a K_{{\bf k}}(t,t_0) n_{{\bf k}}(t_0) \right),	
\eeqa
where we have used the relation $\braket{d_{{\bf k},s}^\dag d_{{\bf k},s'}} = n_{{\bf k},s} \delta_{ss'}$ and $n_{{\bf k}} \equiv {\rm diag}(n_{{\bf k},+},n_{{\bf k},-})$. Similarly, one gets for the spins in the local diagonal basis
\beq
	\mathcal{S}^a_{\bf k}(t) = \frac{1}{2} {\rm Tr} \left( K^\dag_{{\bf k}}(t,t_0) U_{\bf k}(t) \sigma^a U^\dag_{\bf k}(t) K_{{\bf k}}(t,t_0) n_{{\bf k}}(t_0) \right)
\eeq

In order to investigate the spin dynamics, we solve the equation \eq{\ref{eq dot O}} numerically using the Runge-Kutta integrator. For initial conditions, we assume thermal distribution $n_{{\bf k},\pm} = ({\rm exp}(\beta \epsilon_{{\bf k},\pm})+1)^{-1}$ where $\beta$ is the inverse temperature.

For $b_{x,y}(t)$ one could choose any smooth functions with some asymptotic values for $t \rightarrow \pm \infty$. For numerics related reasons, we choose a function such that $\partial_0 b_{x,y} = 0$ at the beginning and the end of the expansion, namely
\beq
b_j(t) = \left\{ 
  \begin{array}{l l}
    1 & \quad t<0 \\
     1 + B_j \sin^2 \left( \frac{\pi}{2}\frac{t}{\tau_j} \right) & \quad 0\leq t \leq \tau_j \\
    1+B_j & \quad t>\tau_j
  \end{array} \right.
\eeq
where $j=x,y$, $B_j$ is the amplitude and $\tau_j$ the duration of the expansion. This gives
\beq
	\partial_0 b_j(t) = B_j \frac{\pi}{2 \tau_j} \sin \left( \pi \frac{t}{\tau_j}\right) \leq B_j \frac{\pi}{2 \tau},
	\label{eq ad criteria num}
\eeq
which can be directly used to evaluate the adiabaticity criteria \eq{\ref{eq ad criteria}}. 

We consider three characteristic cases
\begin{enumerate}[(i)]
\item{massless isotropic ($m=0, b_x = b_y$)}
\item{massless anisotropic ($m=0, b_x \neq b_y$)}
\item{massive isotropic ($m \neq 0, b_x = b_y$)}
\end{enumerate}
The massless isotropic case is trivial, because the field is conformally invariant and there is no associated dynamics of the spins, which we have verified in our simulation. This is in qualitative agreement with the fact, that there are no particle creations in the massless isotropic case \cite{Parker_2009}. The same argument holds for case (ii) for field evaluated at the Dirac points. In order to observe the spin dynamics, one thus has to look in the vicinity, but not directly at the Dirac point.

\begin{center}
	\begin{figure}[t!p]
  	\includegraphics[width=9cm]{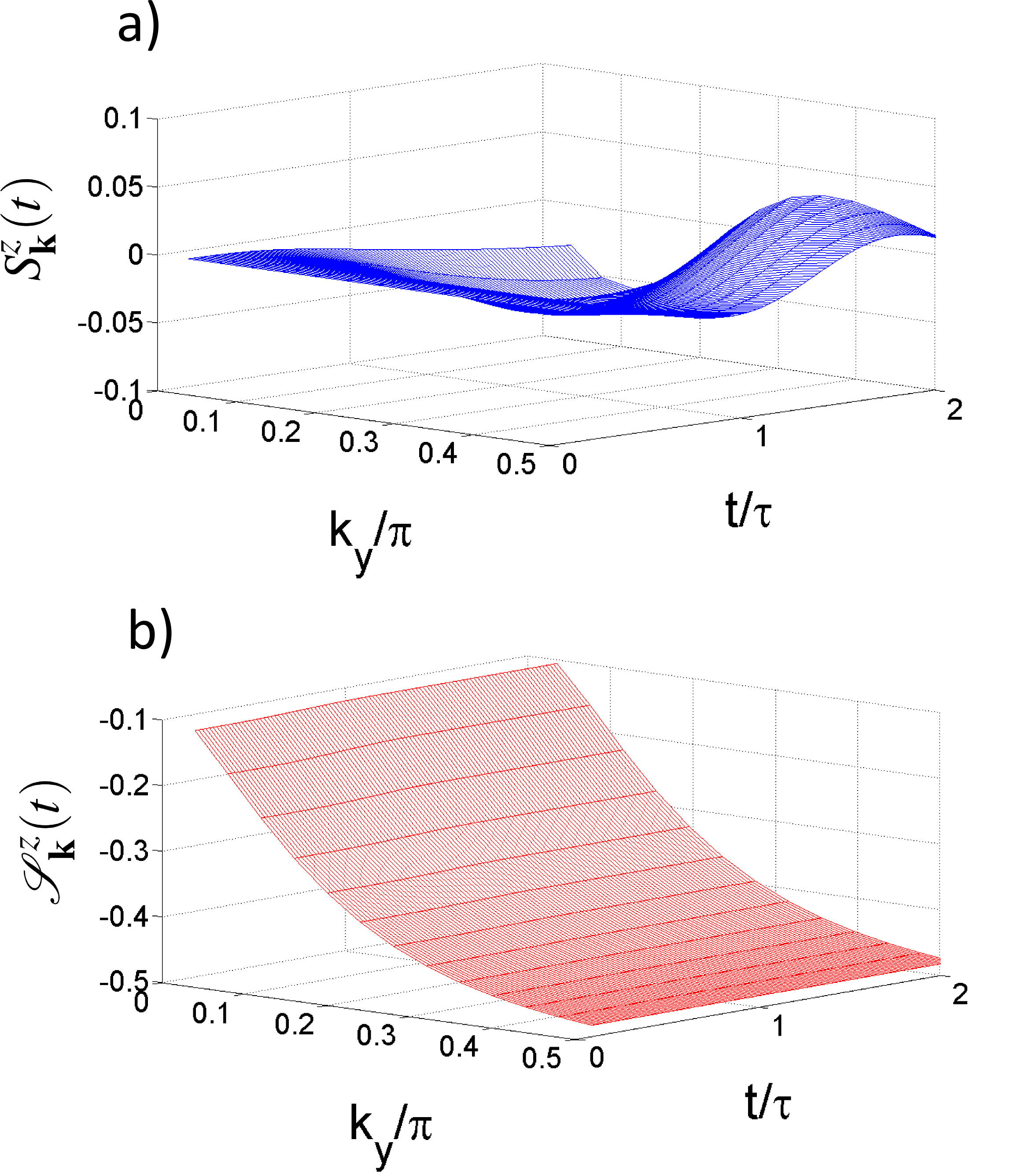} 
  	\caption{(Color online) Massless anisotropic case. z component of the spin in (a) the fermionic basis, $S^z_{\bf k}$ and (b) the diagonal basis, $\mathcal{S}^z_{\bf k}$ vs. time and $k_y$, i.e. the wavevector in the direction which does not undergo the expansion. Used parameters: $m=B_y=0, B_x=1, \tau=5, \beta=5, k_x/\pi=0.02$. See text for details. } 
  	\label{fig spin vs time vs k}
 	\end{figure}
\end{center}

Motivated by the realization in cold atomic experiments, we choose in the following relatively large value of the inverse temperature $\beta=5$.

\begin{center}
	\begin{figure}[t!p]
  	\includegraphics[width=9cm]{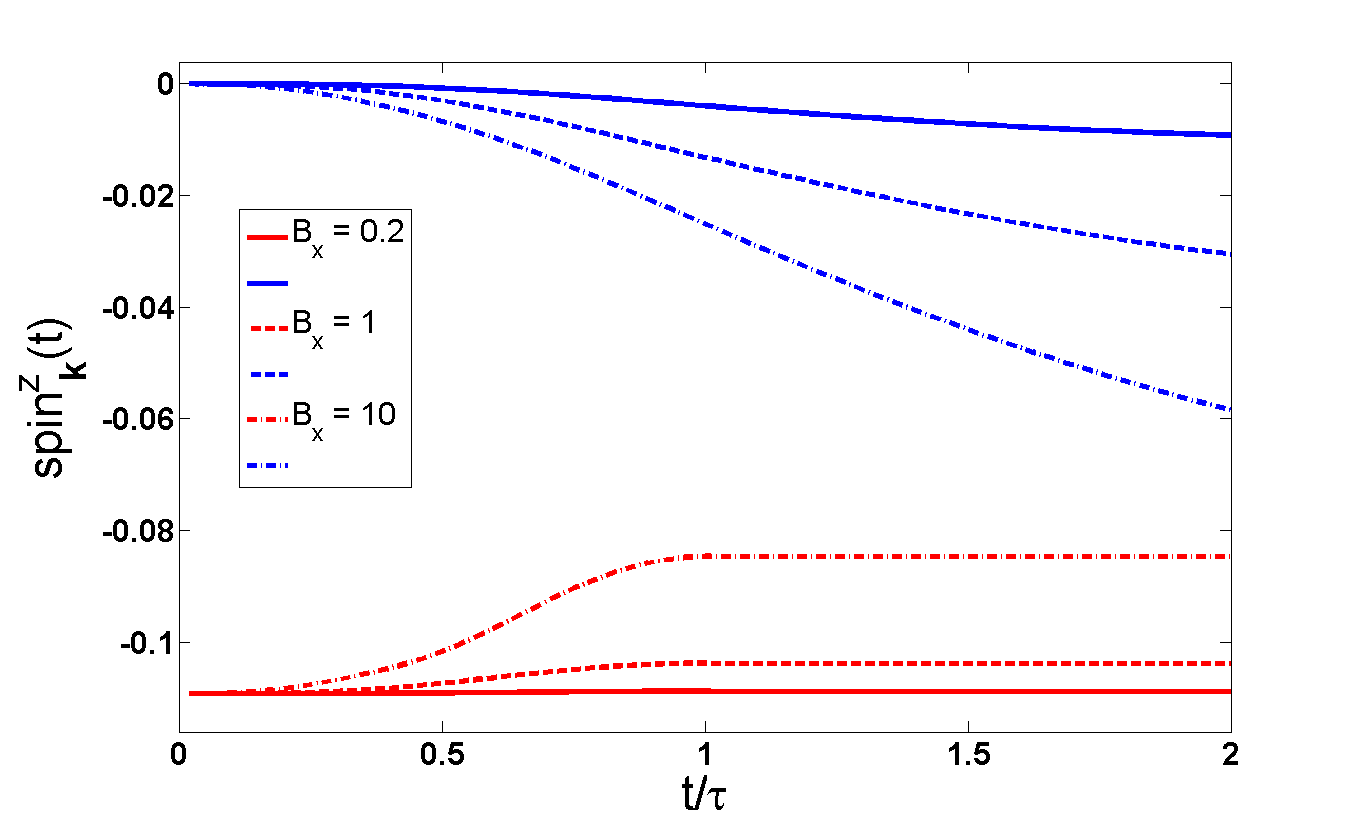} 
  	\caption{(Color online) Massless anisotropic case. z component of the spin in the diagonal basis (red lines), $\mathcal{S}^z_{\bf k}$, and in the fermionic basis (blue lines), $S^z_{\bf k}$, vs. time, evaluated at $(k_x/\pi,k_y/\pi)=(0.02, 0.02)$. Three cases ($B_x = 0.2, 1, 10$) are shown (solid, dashed and dash-dotted lines respectively). Used parameters: $m=B_y=0, \tau=5, \beta=5$. See text for details. } 
  	\label{fig spin vs time ii}
 	\end{figure}
\end{center}

\emph{Massless anisotropic case.} In \fig{\ref{fig spin vs time vs k}} we show the spin dynamics for the massless anisotropic case for different wavevectors $k_y$, i.e. the direction which does not undergo the expansion. The spins in the diagonal basis do not evolve after the expansion is finished ($t/\tau=1$), which is the case for all $\bf k$ (the effect is not clearly visible in \fig{\ref{fig spin vs time vs k}} due to the strong thermal background, see also \fig{\ref{fig spin vs time ii}}). This is in contrast to the spin evolution in the fermionic basis where the spins continue to evolve under the action of the free propagator. This is an important fact with respect to the experimental signatures of the expansion (cf. below).

An example of spin dynamics for different amplitudes of the expansion is shown in \fig{\ref{fig spin vs time ii}}, where we evaluate $S^z_{\bf k}$ and $\mathcal{S}^z_{\bf k}$ in the vicinity of the Dirac point (0,0), namely $(k_x/\pi,k_y/\pi)=(0.02,0.02)$. One can see, as discussed above, that in the diagonal basis, the spin evolution stops when the expansion is finished, as opposed to the fermionic basis. For $B_x = (0.2,1,10)$ using \eq{\ref{eq ad criteria num}} gives $\partial_0 b_x \leq (0.063, 0.31, 3.1)$. According to \eq{\ref{eq ad criteria}}, the cases $B_x = (0.2,1)$ are thus well adiabatic, while $B_x = 10$ is not.

\begin{center}
	\begin{figure}[t!p]
  	\includegraphics[width=9cm]{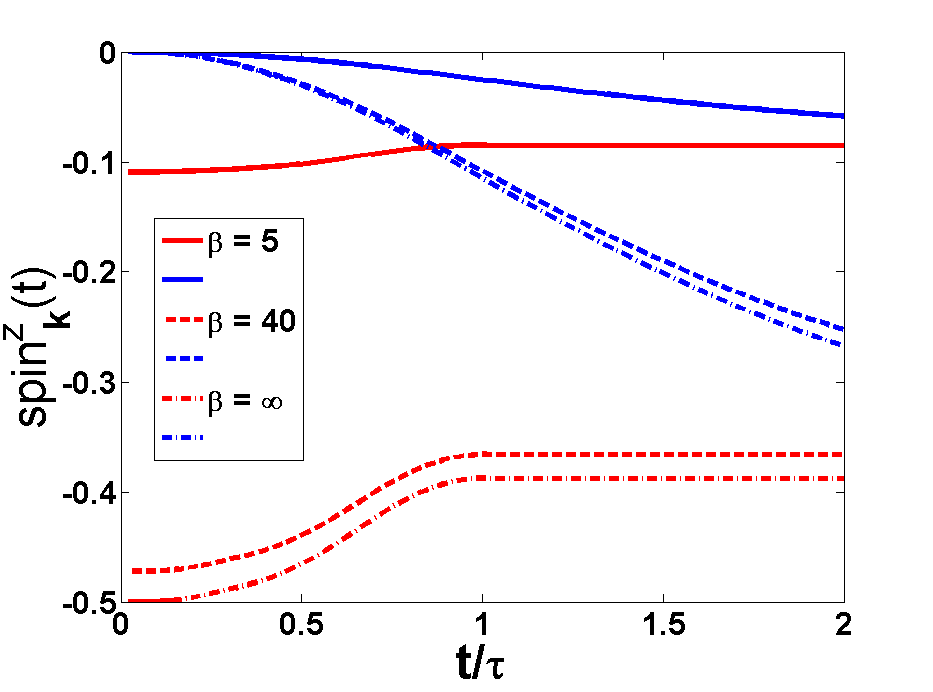} 
  	\caption{(Color online) Massless anisotropic case, temperature dependence. z component of the spin in the diagonal basis (red lines), $\mathcal{S}^z_{\bf k}$, and in the fermionic basis (blue lines), $S^z_{\bf k}$, vs. time, evaluated at $(k_x/\pi,k_y/\pi)=(0.02, 0.02)$. The effect of thermal background for three different temperatures ($\beta = \infty, 40, 5$) is shown (solid, dashed and dash-dotted lines respectively). Used parameters: $m=B_y=0, B_x=1, \tau=5$. See text for details. } 
  	\label{fig spin vs time Beta}
 	\end{figure}
\end{center}

To complete the discussion of the massless anisotropic case, we show in \fig{\ref{fig spin vs time Beta}} the effect of the thermal background for different temperatures ($\beta = \infty, 40, 5$). The qualitative features of the expansion are not affected by the thermal background, however they may be strongly suppressed (e.g. for $\beta=5$).

\begin{center}
	\begin{figure}[t!p]
  	\includegraphics[width=9cm]{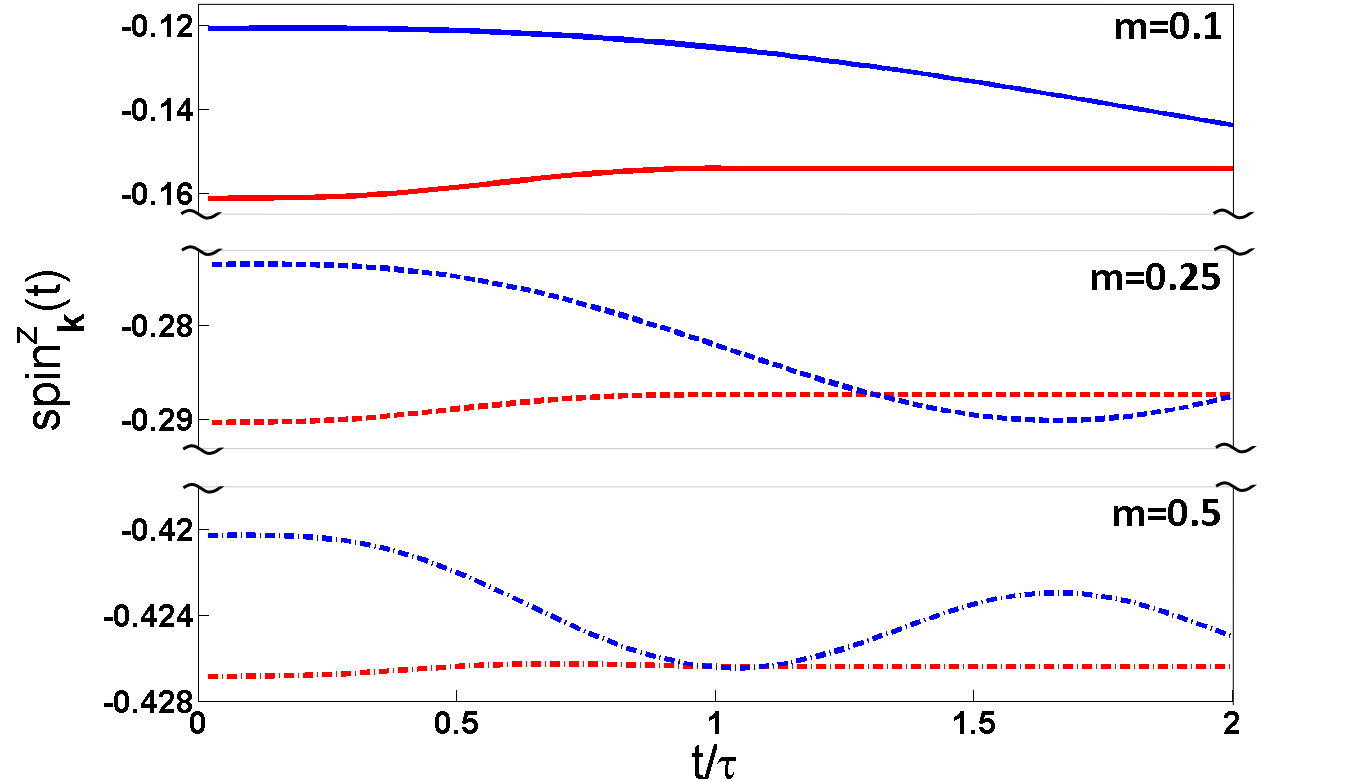} 
  	\caption{(Color online) Massive isotropic case. z component of the spin in the diagonal basis (red lines), $\mathcal{S}^z_{\bf k}$, and in the fermionic basis (blue lines), $S^z_{\bf k}$, vs. time, evaluated at $(k_x/\pi,k_y/\pi)=(0.02, 0.02)$. Three cases ($m = 0.1, 0.25, 0.5$) are shown (solid, dashed and dash-dotted lines respectively). Used parameters: $B_x=B_y=1, \tau=5, \beta=5$. See text for details. } 
  	\label{fig spin vs time iii}
 	\end{figure}
\end{center}

\emph{Massive isotropic case.} Another situation yielding non trivial spin dynamics is an isotropic expansion, but where the field is massive, since the mass term explicitly breaks the conformal invariance. In this case, for small masses the mass term further enforces the restriction to the small $k$ values (cf. the discussion in Sec. \ref{subsec Note on adiabaticity}), $m \gg \frac{k_j}{b_j} \geq k_j$. An example of the spin dynamics for $B_x = B_y = 1$ and $m = (0.1, 0.25, 0.5)$ is shown in \fig{\ref{fig spin vs time iii}}. One can see, that for increasing mass, the expansion has smaller effect on the spin dynamics (decrease in amplitude), i.e. the particle pair creation is suppressed. Another comment is that in the lattice formulation the mass term plays the role of effective magnetic field along $z$ direction, which induces spin precession. This is clearly visible in the fermionic basis, where the precession rate is proportional to the mass (i.e. effective magnetic field), however the amplitude of the precession decreases with increasing mass.

The spin provides an ideal experimental signature of the expansion since it is routinely measured in nowadays cold atomic experiments by probing the populations of atomic levels \cite{Bloch_2008}. Moreover, the time of flight measurements allow to address a specific wavevector $\bf k$ of the Brillouin zone, namely the neighbourhood of the Dirac points.

\subsection{Interactions}

So far, we were considering only non interacting theory. Although non abelian lattice gauge theories are non trivial already at this level, the most interesting physics can be obtained in the presence of interactions. A natural interaction term for spin half fermions in optical lattice is $H_{\rm int} \propto U \sum_{s \neq s'} n_s n_{s'}$, where $n_s = \psi^\dag_s \psi_s$ is the density operator. Once again, one entirely legitimate approach is to consider a Hamiltonian $H = H_{\rm kin} + H_{\rm int}$, with $H_{\rm kin}$ \eq{\ref{eq H kin latt}} and $T \in Gl(2,\mathbb{C})$ as such and study its properties ($H$ could also describe interacting bosons instead of fermions or both bosons and fermions. The interspecies interaction might lead to interesting physical phenomena, such as particle number fractionalization \cite{Ruostekoski_2002, Javanainen_2003, Ruostekoski_2008}). The other approach is to design directly a given field theory. For example, a proposal of simulation of a Thirring model (i.e. 1+1 dimensional field theory) with cold atoms was made \cite{Cirac_2010}, where the interaction term reads $J^\mu J_\mu$ with $J^\mu = \bar{\psi} \gamma^\mu \psi$. In curved spacetime, the replacement $\gamma^\mu \rightarrow \underline{\gamma}^\mu$ makes the interaction term spacetime dependent. One can thus try to modify the proposal \cite{Cirac_2010} in a way that creates the correct interaction term, which might be an interesting test bed situation, since as one dimensional theory, the massless Thirring model is soluble also in curved spacetime \cite{Birell_1978}.

\section{Conclusion}
\label{sec Conclusion}

In this article we have shown how to map the continuous Dirac fields in curved background spacetime to the Fermi-Hubbard model with general nonunitary tunnelings both in the relativistic and non-relativistic cases. Next, we have demonstrated the mapping on the example of an expanding FLRW universe in 2+1 dimensions. Motivated by the experimental feasibility of such Hamiltonian in cold atomic experiments with laser assisted tunnelings, we could explicitly demonstrate the effect of time dependent non unitary tunnelings on the spin dynamics. We found, that the dynamics of the spin (representing the particle mode occupation) shares the same qualitative features as those predicted by the quantum field theory in FLRW spacetime, namely the dependence on mass and no particle creation in the massless isotropic (conformally invariant) case.

\subsection*{Acknowledgements}

We would like to thank J. Baez, T. Brauner, A. Kempf, T. Lappi, V. Scarani and M.C. Tan for useful discussions. J.M. acknowledges the support by the National Research Foundation and the Ministry of Education, Singapore. The Centre for Quantum Technologies is a Research Centre of Excellence funded by the Ministry of Education and the National Research Foundation of Singapore.

\bibliography{CurvedDirac_paper_arXiv_v2}
\bibliographystyle{prsty}

\section*{Appendix}

\subsection{Nonrelativistic limit}

Next, we discuss a non relativistic limit of the Dirac equation. After all, the kinetic part of the usual Hubbard model for electrons in tight binding approximation (\eq{\ref{eq H kin latt}} with $T \propto \mathds{1}$) is obtained from the non relativistic quantum mechanical Hamiltonian $H \propto {\bf p}^2/(2m)$. It is thus interesting to see, how similar derivation works for fermions in curved spacetime background. We should use a systematic method, known as Foldy-Wouthouysen transformation \cite{Bjorken_1964} (also used in the context of quantum fields in curved spacetimes \cite{Bakke_2008}), which perturbatively decouples the electron and positron modes. One can derive the Dirac equation from \eq{\ref{eq L}}
\beq
	(i\underline{\gamma}^\mu D_\mu - m) \psi(x) = 0,
\eeq
which can be rewritten as Schr\"{o}dinger equation
\beq
	i\partial_0 \psi(x) = \mathcal{H}_D \psi(x),
\eeq
where
\beq
	\mathcal{H}_D = (\underline{\gamma}^0)^{-1} (m - i\underline{\gamma}^k D_k) + i\Gamma_0
	\label{eq HD dens}
\eeq
is the Dirac Hamiltonian. The non relativistic limit can be obtained from the Dirac Hamiltonian of the form
\beq
	\mathcal{H}_D = \gamma^0 m + \mathcal{E} + O,
	\label{eq HD OE}
\eeq
where $\mathcal{E}$ and $O$ are even and odd operator, defined by the property $\left[ \gamma^0, \mathcal{E} \right] = 0$ and $\{\gamma^0, O\} = 0$ and $\gamma^0$ is in the Dirac representation. The lowest order expression for the nonrelativistic Hamiltonian is
\beq
	\mathcal{H}_P = \gamma^0 m + \mathcal{E} + \frac{1}{2m} \gamma^0 O^2,
\eeq
where the subscript $P$ stands for the Pauli Hamiltonian. One can identify $\gamma^0, \mathcal{E}$ and $O$ by comparing \eq{\ref{eq HD OE}} with \eq{\ref{eq HD dens}}. In the most general case it yields rather lengthy expressions. In order to proceed with the calculation, we will thus consider a simple, yet non-trivial scenario with a static diagonal metric of the form
\beq
	g = \left( \begin{array}{cc}
	1 & 0 \\
	0 & h
	\end{array}\right),
	\label{eq g stat}
\eeq
and $h = {\rm diag}(h_{ii}(x^k))$, where $i = 1..d-1$ and the diagonal terms depend only on spatial coordinates $x^k$. First thing we note, is that in this case, the vielbein fields are also diagonal, $e^\mu_\alpha = 0$ for $\mu \neq \alpha$. In particular $e^0_0 = 1$ implying $(\underline{\gamma}^0)^{-1} = \underline{\gamma}^0 = \gamma^0$. Also, $\Gamma_0 = 0$. We then obtain for the Dirac Hamiltonian
\beq
	\mathcal{H}_D = \gamma^0 m - i \gamma^0 \underline{\gamma}^k D_k = \gamma^0 m + O,
\eeq
since the term $\gamma^0 \underline{\gamma}^k D_k$ is odd for the metric considered. We then obtain for the Pauli Hamiltonian density
\beq
	\mathcal{H}_P = \gamma^0 m - \frac{1}{2m} \gamma^0 (\gamma^0 \underline{\gamma}^k D_k) (\gamma^0 \underline{\gamma}^j D_j).
\eeq
The total Hamiltonian, expressed in terms of field variables, then reads
\beq
	H_P = \frac{1}{2}\left[ (\psi, \mathcal{H}_P \psi) + (\mathcal{H}_P \psi, \psi) \right].
	\label{eq H Pauli def}
\eeq
The scalar product can be evaluated by integrating per parts in curved spacetime. The reason why one wants to do that is to obtain terms of type $(D \psi^\dag) (D \psi)$ rather than $\psi^\dag D^2 \psi$, since the former can be mapped to a Hubbard model with only nearest neighbor hopping. Evaluating \eq{\ref{eq H Pauli def}}, we get
\beqa
	&& H_P = \frac{1}{2}\frac{1}{2m} \int {\rm d}^{d-1} x \sqrt{g} \bar{\psi} \underline{\gamma}^k \underline{\gamma}^j D_k D_j \psi + {\rm h.c.} \nonumber \\
	&&  + \int {\rm d}^{d-1} x \sqrt{g} m \bar{\psi} \psi 	
\eeqa
At this point $\psi$ is still $2^{\left[d/2\right]}$ component spinor, where $\left[n\right]$ is the integer part of $n$. By construction, the Hamiltonian $H_P$ contains only even operators and we can thus split the spinor into two parts, say $\psi = (\chi, \varphi)$. Each of the spinors $\chi, \varphi$ has $2^{\left[d/2\right]-1}$ components, which will have independent dynamics. In case of the diagonal static metric and $d=4$, we find
\beq
	\underline{\gamma}^k \underline{\gamma}^j D_k D_j = - \left( \begin{array}{cc}
		e^k_k e^j_j \sigma^k \sigma^j \nabla_k \nabla_j & 0 \\
		0 & e^k_k e^j_j \sigma^k \sigma^j \nabla_k \nabla_j
	\end{array} \right),
\eeq
where $\nabla_k = \partial_k - \tilde{\Gamma}_k$, $\tilde{\Gamma}_k = \left. -1/4 \sigma^j \sigma^l e^\nu_j({\bf x}) (\nabla_k e_{l \nu}({\bf x})) \right|_{j<l}$. Lets write the Pauli Hamiltonian for one of the spinor components, say $\chi$, which we write as
\beqa
	&& \int {\rm d}^{3} x \sqrt{g} \chi^\dag \underline{\sigma}^k \underline{\sigma}^j \nabla_k \nabla_j \chi = \nonumber \\
	&& \int {\rm d}^{3} \chi^\dag f_{ii} \nabla_i \nabla_i \chi + \int {\rm d}^{d-1} \left. \chi^\dag f_{kj} \sigma^k \sigma^j [\nabla_k, \nabla_j] \right|_{k<j} \chi,
\eeqa
where $f_{kj} = \sqrt{g} e^k_k e^j_j$. The commutator in the second term is familiar from non-abelian gauge theories and we have $[\nabla_k, \nabla_j] = \partial_{\left[ j \right.} \tilde{\Gamma}_{\left. k \right]} - \left[ \tilde{\Gamma}_k, \tilde{\Gamma}_j \right]$, which acts locally on the spinor $\chi$. The first term can be integrated per parts to yield (using $\tilde{\Gamma}_i^\dag = -\tilde{\Gamma}_i$)
\beqa
	&& \int {\rm d}^{3} x \chi^\dag f_{ii} \nabla_i \nabla_i \chi = \nonumber \\
	&& - \int {\rm d}^{3} x \left\{ f_{ii} (\nabla_i \chi)^\dag (\nabla_i \chi) -  (\partial_i f_{ii}) \chi^\dag (\nabla_i \chi) \right\}.
\eeqa
We thus write the Pauli Hamiltonian as
\begin{widetext}
\beq 
	H_P = \int {\rm d}^{3} x \frac{1}{2}\frac{1}{2m} \left[ f_{ii} (\nabla_i \chi)^\dag (\nabla_i \chi) -  (\partial_i f_{ii}) \chi^\dag (\nabla_i \chi) - \left. \chi^\dag f_{kj} \sigma^k \sigma^j [\nabla_k, \nabla_j] \right|_{k<j} \chi, \right] + {\rm h.c.} + \sqrt{g} m \chi^\dag \chi.
	\label{eq H Pauli}
\eeq
\end{widetext}
We are now in the position to discretize the Pauli Hamiltonian, which is to follow exactly the same steps as in the case of Dirac Hamiltonian. Using again the prescription \eq{\ref{eq psi xt psi x}}, which now takes a simple form $\chi(x) = \sqrt{2}\sqrt{g}^{-\frac{1}{2}} X({\bf x})$, we arrive at a Hamiltonian, which can be formally written as \eq{\ref{eq H t}}, where we have to replace $\Psi \rightarrow X$ and the matrices $T, V$ now depends only on spatial coordinates $\bf x$ and read
\beqa
	&& T({\bf x},{\bf x} + a_i) = -\frac{\sqrt{g}^{-1}}{m} (f_{ii} + \frac{1}{2}\partial_i f_{ii}) P({\bf x},{\bf x} + a_i) \nonumber \\
	&& V({\bf x}) = 2m + \frac{\sqrt{g}^{-1}}{m} \left[ f_{ii} + f^{-}_{ii} - (\partial_i f_{ii}) \right. \nonumber \\
	&&  \phantom{V({\bf x}) =} + \left. \left( \frac{1}{2}\sqrt{g} \left. \underline{\sigma}^k \underline{\sigma}^j \left[\nabla_k, \nabla_j \right] \right|_{k<j}  + {\rm h.c.} \right) \right],	
\eeqa
where $f^{-}_{ii} = f_{ii}({\bf x} - a_i )$ and we have to replace $\Gamma \rightarrow \tilde{\Gamma}$ in the definition of the parallel propagator \eq{\ref{eq par prop}}.

It is interesting to notice, that in the case of relativistic Dirac Hamiltonian taking static diagonal metric \eq{\ref{eq g stat}}, $\underline{\gamma}$ are unitary, $\Gamma_k$ are antihermitian, $\Gamma_k^\dag = -\Gamma_k$ and the parallel propagators \eq{\ref{eq par prop}} become unitary. We thus have $T$ which is also unitary, contrary to the Pauli Hamiltonian case.

\end{document}